# Time delayed processes in physics , biophysics and archaeology

Magdalena Anna Pelc

Institute of Physics
Maria Curie-Sklodowska University, Lublin, Poland

Abstract

The motion of "particles". where the particles: electrons, ions in microtubules or migrated peoples can be described as the superposition of diffusion and ordered waves. In this paper it is shown that the master equation for transport processes can be formulated as the time delayed hyperbolic partial equation. The equation describes the processes with memory For characteristic times shorter than the relaxation time  the master  equation is the generalized Klein - Gordon equation.

Introduction

A key quantity in the description of the transport phenomena is the relaxation time of the corresponding dissipative processes. In the time dealyed description of the transport phenomena we argue that the generalized $q(\vec{r},t)$ current of the "partons", for example electrons, or migrated peoples fulfills the relaticent

$$q(t+\tau) = q(t) + \tau \frac{\partial q}{\partial t}...$$

(I.1)

Following formula (I.1) and conservation laws for transport processes the master equation for transport phenomena can be obtained:

$$\tau \frac{\partial^2 f(\vec{r},t)}{\partial t^2} + \frac{\partial f(\vec{r},t)}{\partial t} = D\nabla^2 f(\vec{r},t) \qquad (I.2)$$

In this paper the master equation (I.2) will be used to the description of the time delayed phenomena in physics, biology and archaeology.

# Chapter 1

# Overview of the research

## 1.1 Hyperbolic versus parabolic

In the description of the evolution of any physical system, it is mandatory to evaluate, as accurately as possible, the order of magnitude of different characteristic time scales, since their relationship with the time scale of observation (the time during which we assume our description of the system to be valid) will determine, along with the relevant equations, the evolution pattern. Take a forced damped harmonic oscillator and consider its motion on a time scale much larger than both the damping time and the period of the forced oscillation. Then, what one observes is just a harmonic motion. Had we observed the system on a time scale of the order of (or smaller) than the damping time, the transient regime would have become apparent. This is rather general and of a very relevant interest when dealing with dissipative systems. It is our purpose here, by means of examples and arguments related to a wide class of phenomena, to emphasize the convenience of resorting to hyperbolic theories when dissipative processes, either outside the steady-state regime or when the observation time is of the order or shorter than some characteristic time of the system, are under consideration. Furthermore, as it will be mentioned below, transient phenomena may affect the way in which the system leaves the equilibrium, thereby affecting the future of the system even for time scales much larger than the relaxation time.

Parabolic theories of dissipative phenomena have long and a venerable history and proved very useful especially in the steady-state regime [1.1]. They exhibit however some undesirable features, such as acausality (see e.g., [1.2], [1.3]), that prompted the formulation of hyperbolic theories of dissipation to get rid of them. This was achieved at the price of extending the set of field variables by including the dissipative fluxes (heat current, non-equilibrium stresses and so on) at the same footing as the classical ones (energy densities, equilibrium pressures, etc), thereby giving rise to a set of more physically satisfactory (as they much better conform with experiments) but involved theories from the mathematical point of view. These theories have the additional advantage of being backed by statistical fluctuation theory, kinetic theory of gases (Grad's 13-moment approximation), information theory and correlated random walks (at least in the version of Jou *et al.*) [1.3].

A key quantity in these theories is the relaxation time $\tau$ of the corresponding dissipative process. This positive-definite quantity has a distinct physical meaning, namely the time taken by the system to return spontaneously to the steady state (whether of thermodynamic equilibrium or not) after it has been suddenly removed from it. It is, however, connected to the mean collision time $t_c$ of the particles responsible for the dissipative process It is therefore appropriate to interpret the relaxation time as the time taken by the corresponding dissipative flow to relax to its steady value. Thus, it is well known that the classical Fourier law for heat current, [1.3]

$$\vec{q} = -\kappa \nabla T \tag{1.1}$$

with $\kappa$ the heat conductivity of the fluid, leads to a parabolic equation for temperature (diffusion equation)

$$\frac{\partial T}{\partial t} = \chi \nabla^2 T, \qquad \chi = \frac{\kappa}{\rho C_p} \tag{1.2}$$

(where $\chi$, $\rho$ and $C_p$ are diffusivity, density and specific heat at constant pressure, respectively), which does not forecast propagation of perturbations along characteristic causal light-cones That is to say, perturbations propagate with infinite speed. This non-causal behavior is easily visualized by taking a look at the thermal conduction in an infinite one dimensional medium. Assuming that the temperature of the line is zero for $t < 0$, and putting a heat source at $x = x_0$ when $t = 0$, the temperature profile for $t > 0$ is given by

$$T \sim \frac{1}{\sqrt{t}} \exp\left[-\frac{(x-x_0)^2}{t}\right] \tag{1.3}$$

implying that for $t = 0 \Rightarrow T = \delta(x - x_0)$, and for $t = t_1 > 0 \Rightarrow T \neq 0$. In other words, the presence of a heat source at $x_0$ is instantaneously felt by all observers on the line, no matter how far away from $x_0$ they happen to be. The origin of this behavior can be traced to the parabolic character of Fourier's law, which implies that the heat flow starts (vanishes) simultaneously with the appearance (disappearance) of a temperature gradient. Although $\tau$ is very small for phonon-electron, and phonon-phonon interaction at room temperature ($10^{-11}$ and $10^{-13}$ seconds, respectively), neglecting it is the source of difficulties, and in some cases a bad approximation as for example in superfluid Helium, and degenerate stars where thermal conduction is dominated by electrons -see [1.3], for further examples.

In order to overcome this problem Cattaneo and (independently) Vernotte by using the relaxation time approximation to Boltzmann equation for a simple gas derived a generalization of Fourier's law, namely [1.3]

$$\tau \frac{\partial \vec{q}}{\partial t} + \vec{q} = -\kappa \nabla T \qquad (1.4)$$

This expression (known as Cattaneo-Vernotte's equation) leads to a hyperbolic equation for the temperature (Heaviside equation) which describes the propagation of thermal signals with a finite speed

$$v = \sqrt{\chi/\tau} \qquad (1.5)$$

This diverges only if the unphysical assumption of setting $\tau$ to zero is made.

It is worth mentioning that a simple random walk analysis of transport processes naturally leads to Heaviside equation, not to the diffusion equation -see e.g. [1.4]. Again, the latter is obtained only if one neglects the second derivative term.

It is instructive to write (1.4) in the equivalent integral form

$$\vec{q} = -\frac{\chi}{\tau} \int_{-\infty}^{t} \exp\left[-\frac{(t-t')}{\tau}\right] \cdot \nabla T(x',t') dt' \qquad (1.6)$$

which is a particular case of the more general expression

$$\vec{q} = -\int_{-\infty}^{t} Q(t-t') \nabla T(x',t') dt' \qquad (1.7)$$

The physical meaning of the kernel $Q(t-t'')$ becomes obvious by observing that

$$\text{for} \quad Q = \kappa \delta(t-t') \rightarrow \vec{q} = -\kappa \nabla T \qquad \text{(Fourier)} \qquad (1.8)$$

$$\text{for} \quad Q = \text{constant} \rightarrow \frac{\partial^2 T}{\partial t^2} = \chi \nabla^2 T \qquad \text{(wave motion)}$$

i.e., $Q$ describes the thermal memory of the material by assigning different weights to temperature gradients at different moments in the past. The Fourier law corresponds to a zero-memory material (the only relevant temperature gradient is the "last" one, i.e., the one simultaneous with the appearance of *q)*. By contrast the infinite memory case (with *Q* = constant) leads to an undamped wave. Somewhere in the middle is the Cattaneo-Vernotte equation, for which all temperature gradients contribute to *q,* but their relevance goes down as we move to the past.

From these comments it should be clear that different classes of dissipative systems may be described by different kernels. The one corresponding to (1.4) being suitable for the description of a restricted subclass of phenomena.

Obviously, when studying transient regimes, i.e., the evolution from a steady-state situation to a new one, τ cannot be neglected. In fact, leaving aside that parabolic theories are necessarily non-causal, it is obvious that whenever the time scale of the problem under consideration becomes of the order of (or smaller) than the relaxation time, the latter cannot be ignored. It is common sense what is at stake here: neglecting the relaxation time amounts -in this situation- to disregarding the whole problem under consideration. According to a basic assumption underlying the disposal of hyperbolic dissipative theories, dissipative processes with relaxation times comparable to the characteristic time of the system are out of the hydrodynamic regime. However, the concept of hydrodynamic regime involves the ratio between the mean free path of fluid particles and the characteristic length of the system. When this ratio is lower that unity, the fluid is within the hydrodynamic regime. When it is larger than unity, the regime becomes Knudsen's. In the latter case the fluid is no longer a continuum and even hyperbolic theories cease to be realiable.

Therefore that assumption can be valid only if the particles making up the fluid are the same ones that transport the heat. However, this is never the case. Specifically, for a neutron star, r is of the order of the scattering time between electrons (which carry the heat) but this fact is not an obstacle (no matter how large the mean free path of these electrons may be) to consider the neutron star as formed by a Fermi fluid of degenerate neutrons. The same is true for the second sound in superfluid Helium and solids, and for almost any ordinary fluid. In brief, the hydrodynamic regime refers to fluid particles that not necessarily (and as a matter of fact, almost never) transport the heat. Therefore large relaxation times (large mean free paths of particles involved in heat transport) does not imply a departure from the hydrodynamic regime, but it is usually overlooked).

However, even in the case that the particles making up the fluid are responsible of the dissipative process, it is not "always" valid to take for granted that $\tau$ and $t_c$ are of the same order or what comes to the same that the dimensionless quantity $\Gamma = (\tau c_s/L)^2$ is negligible in all instances – here $c_s$ stands for the adiabatic speed of sound in the fluid under consideration and $L$ the characteristic length of the system. That assumption would be right if $\tau$ were always comparable to $t_c$ and $L$ always "large", but there are, however, important situations in which $\tau >> t_c$ and $L$ "small" although still large enough to justify a macroscopic description. For tiny semiconductor pieces of about $10^{-4}$ cm in size, used in common electronic devices submitted to high electric fields, the above dimensionless combination (with $\tau \sim 10^{-10}$ sec, $c_s \sim 10^7$ cm/sec [1.3]) can easily be of the order of unity. In ultrasound propagation as well as light-scattering experiments in gases and neutron-scattering in liquids the relevant length is no longer the system size, but the wave length A which is usually much smaller than $L$ Because of this, hyperbolic theories may bear some importance in the study of nanoparticles and quantum dots. Likewise m polymeric fluids relaxation times are related to the internal configurational degrees of freedom and so much longer than $t_c$ (in fact they are in the range of the minutes), and $c_a \sim 10^5$ cm/sec, thereby $\Gamma \sim 1$. In the degenerate core of aged stars the thermal relaxation time can be as high as l second [1.2]. Assuming the radius of the core of about $10^{-2}$ times the solar radius, one has $\Gamma \sim 1$ again. Fully ionized plasmas exhibit a collisionless regime (Vlasov regime) for which the parabolic hydrodynamics predicts a plasmon dispersion relation at variance with the microscopic results; the latter agree, however, with the hyperbolic hydrodynamic approach. Think for instance of some syrup fluid flowing under a imposed shear stress, and imagine that the shear is suddenly switched off. This liquid will come to a halt only after a much longer time ($\tau$) than the collision time between its constituent particles has elapsed [1.2].

## 1.2 High Energy Density Systems

The increase in light intensity available in the laboratory over the previous 20 years has been astounding. Laser peak power has climbed from gigawatts to petawatts in this time span, and accessible focused intensity has increased by at least seven orders of magnitude. Such a dramatic increase in light brightness has accessed an entirely new set of phenomena. High repetition rate table top lasers can routinely produce intensity in excess of $10^{19}$W/cm$^2$, and intensities of up to $10^{20}$W/cm$^2$ are possible with the latest petawatt class systems. Light-matter interactions with single atoms are strongly non-perturbative and electron energies are relativistic. The intrinsic energy density of these focused pulses is very high, exceeding a gigajoule per $cm^3$. The interactions of such intense light with matter lead to dramatic effects, such as high temperature plasma creation, bright X-ray pulse generation, fusion plasma production, relativistic particle acceleration, and highly charged ion production [1.5].

Such exotic laser-matter interactions have led to an interesting set of applications in high field science, and high energy density physics (HED physics). These applications span basic science and extend into unexpected new areas such as fusion energy development and astrophysics. In this paper some of these new applications will be reviewed. The topics covered here do not represent a comprehensive list of applications made possible with high intensity short pulse lasers, but they do give a flavor of the diverse areas affected by the latest laser technology. Most of the applications discussed here are based on recent experiments using lasers with peak power of 5-100 TW. Many important leaps in laser technology have driven the rapid advances in HED and high field science over the last 15 years. The enabling advancement for this technological progress was the invention of chirped pulse amplification (CPA) lasers.

The physics accessible with this class of lasers is quite extreme. The science applications made possible with these extremes can be simply classified into two categories. First, by temporally compressing the pulses and focusing to spots of a few wavelengths in diameter, these lasers concentrate energy in a very small volume. A multi-TW laser focused to a few microns has an intrinsic energy density of over $10^9$ $J/cm^3$. This corresponds to about l0 keV of energy per atom in a material at solid density. As a result, quite high temperatures can be obtained. Such energy density corresponds to pressure in excess of l0 Gbar. Many applications of ultraintense lasers stem from the ability to concentrate energy to high energy-density which can lead to quite extreme states of matter.

The second class of applications arises from the high field strengths associated with a very intense laser pulse. At an intensity of over $10^{18}$ $W/cm^2$, an intensity quite easily achievable with modern table top terawatt lasers, the electric field of the laser exceeds ten atomic units, over $10^{10}$ V/cm. Consequently, the strong field rapidly ionizes the atoms and molecules during a few laser cycles. At intensity approaching $10^{21}$ $W/cm^2$, the current state of the art with petawatt class lasers, the electric field is comparable to the field felt by a K-shell electron in mid Z elements such as argon. The magnetic field is nearly 1000 T. An electron quivering in such a strong electric field will be accelerated to many MeV of energy in a single optical cycle. Such charged particles will experience a very strong forward directed force resulting from the Lorentz force [1.5].

High peak power femtosecond lasers are unique in their ability to concentrate energy in a small volume. A dramatic consequence of this concentration of energy is the ability to create matter at high temperature and pressure. Matter with temperature and density near the center of dense stars can be created in the laboratory with the latest high intensity lasers. For example, solid density matter can be heated to temperature of over $10^6$ C > 1 keV. Under these conditions, the particle pressure inside the sample is over 1 billion atmospheres, far higher than any other pressure found naturally on or in the earth and approaches pressures created in nuclear weapons and inertial confinement fusion implosions.

Study of the properties of matter at these extreme conditions, namely solid density or higher in the temperature range of 1 – 1000 eV, is crucial to understanding many diverse phenomenon, such as the structure of planetary and stellar interiors or how controlled nuclear fusion implosions (inertial confinement fusion or ICF) evolve. Yet, despite the wide technological and astrophysical applications, a true, complete understanding of matter in this regime is not in hand. A large obstacle is posed by the fact that theoretical models of this kind of matter are difficult to formulate. While the atoms in these warm and hot dense plasmas are strongly ionized, the very strong coupling of the plasma, and continuum lowering in the plasma dramatically complicates traditional plasma models which depend on two body collision kinetics. Even the question of whether electrons in this state are free or bound is not as clear cut as it is in a diffuse plasma.

## 1.3. High Energy Density Heat Transport and Special Relativity Theory

In the interaction of laser beam with thin solid target, fast ion beams and electrons are generated with large density. When the energy of laser beam is highly enough the ion and electron velocities are relativistic and transport of thermal energy must be described by relativistic equations. In the paper [1.6] we develop the description of the heat transport in Minkowski spacetime. In the context special relativity theory we investigate the Fourier equation and hyperbolic diffusion equation. We calculate the speeds of the heat diffusion in Fourier approximation and show that for high energy laser beam the heat diffusion exceeds the light velocity. We show that this results breaks the causality of the thermal phenomena in Minkowski spacetime. The same phenomena we describe in the framework of hyperbolic heat diffusion equation and show that in that case speed of diffusion is always smaller than light velocity.

We may use the concept that the speed of light *in vacuo* provides an upper limit on the speed with which a signal can travel between two events to establish whether or not any two events could be connected. In the interest of simplicity we shall work with one space dimension $x_1 = x$ and the time dimension $x_o = ct$ of the Minkowski spacetime.

Let us consider events (1) and (2): their Minkowski interval *Δs* satisfies the relationship:

$$\Delta s^2 = c^2 \Delta t^2 - \Delta x^2 \qquad (1.9)$$

Without loss of generality we take Event 1 to be at $x = 0$, $t = 0$. Then Event 2 can be only related to Event 1 if it is possible for a signal traveling at the speed of light, to connect them. Event 2 is at (*Δx, cΔt*), its relationship to Event 1 depends on whether *Δs* > 0, = 0, or < 0.

We may summarize these three possibilities as follows:

Case A   *timelike interval*, $|\Delta x_A| < c\Delta t$, or $\Delta s^2 > 0$. Event 2 can be related to Event 1, events 1 and 2 can be in causal relation.

Case B   *lightlike interval*, $|\Delta x_B| = c\Delta t$, or $\Delta s^2 = 0$. Event 2 can only be related to Event 1 by a light signal.

Case C   *spacelike interval* $|\Delta x_A| > c\Delta t$, or $\Delta s^2 < 0$. Event 2 cannot be related to Event 1, for in that case $\upsilon > c$.

Now let us consider the case C in more details. At first sight it seems that in case C we can find out the reference frame in which two Events $c^>$ and $c^<$ always fulfils the relations $t_{c^>} - t_{c^<} > 0$. but it is not true. For lest us choose the inertial frame U' in which $t_{c^>} - t_{c^<} > 0$. In reference frame U which is moving with speed *V* relative to U', where

$$V = c\frac{c(t'_{c_>} - t'_{c_<})}{x'_{c_<} - x'_{c_>}} \tag{1.10}$$

Speed $V < c$ for

$$\left|\frac{c(t'_{c_>} - t'_{c_<})}{x'_{c_<} - x'_{c_>}}\right| < 1 \tag{1.11}$$

Let us calculate $t_{c_>} - t_{c_<}$ in the reference frame U

$$
\begin{aligned}
t_{c_>} - t_{c_<} &= \frac{1}{\sqrt{1-\frac{V^2}{c^2}}}\left[\frac{V}{c^2}(x'_{c_>} - x'_{c_<}) + (t'_{c_>} - t'_{c_<})\right] = \\
&\frac{1}{\sqrt{1-\frac{V^2}{c^2}}}\left[\frac{t'_{c_>} - t'_{c_<}}{x'_{c_>} - x'_{c_<}}(x'_{c_>} - x'_{c_<}) + (t'_{c_>} - t'_{c_<})\right] = 0
\end{aligned}
\tag{1.12}
$$

For the greater $V$ we will have $t_{c_>} - t_{c_<} < 0$. It means that for the spacelike intervals the sign of $t_{c_>} - t_{c_<}$ depends on the speed $V$, i.e. causality relation for spacelike events is not valid.

## 1.4. Fourier diffusion equation and special relativity theory

In paper [1.7] the speed of the diffusion signals was calculated

$$\upsilon = \sqrt{2D\omega} \tag{1.13}$$

where

$$D = \frac{\hbar}{m} \tag{1.14}$$

and $\omega$ is the angular frequency of the laser pulses. Considering formula (1.13) and (1.14) one obtains

$$\upsilon = c\sqrt{2\frac{\hbar\omega}{mc^2}} \tag{1.15}$$

and $\upsilon \geq c$ for $\hbar\omega \geq mc^2$.

From formula (1.15) we conclude that for $\hbar\omega > mc^2$ the Fourier diffusion equation is in contradiction with special relativity theory and breaks the causality in transport phenomena.

In monograph [1.9] the hyperbolic model of the heat transport phenomena was formulated. It was shown that the description of the ultrashort thermal energy transport needs the hyperbolic diffusion equation (one dimension transport)

$$\tau \frac{\partial^2 T}{\partial t^2} + \frac{\partial T}{\partial t} = D \frac{\partial^2 T}{\partial x^2} \qquad (1.16)$$

In the equation (1.16) $\tau = \dfrac{\hbar}{m\alpha^2 c^2}$ is the relaxation time, $m$ = mass of the heat carrier, $\alpha$ is the coupling constant and $c$ is the light speed in vacuum, $T(x,t)$ is the temperature field and $D = \hbar/m$.

In paper [1.7] the speed of the thermal propagation $v$ was calculated

$$v = \frac{2\hbar}{m}\sqrt{-\frac{m}{2\hbar}\tau\omega^2 + \frac{m\omega}{2\hbar}(1+\tau^2\omega^2)^{\frac{1}{2}}} \qquad (1.17)$$

Considering that $\tau = \dfrac{\hbar}{m\alpha^2 c^2}$ formula (1.17) can be written as

$$v = \frac{2\hbar}{m}\sqrt{-\frac{m}{2\hbar}\frac{\hbar\omega^2}{mc^2\alpha^2} + \frac{m\omega}{2\hbar}(1+\frac{\hbar^2\omega^2}{m^2c^4\alpha^4})^{\frac{1}{2}}} \qquad (1.18)$$

For

$$\frac{\hbar\omega}{mc^2\alpha^2} < 1, \quad \frac{\hbar\omega^2}{mc^2} < 1 \qquad (1.19)$$

one obtains from formula (1.29)

$$v = \sqrt{\frac{2\hbar}{m}\omega} \qquad (1.20)$$

Formally formula (1.20) is the same as formula (1.15) but considering inequality (1.18) we obtain

$$v = \sqrt{\frac{2\hbar\omega}{m}} = \sqrt{2}\alpha c < c \qquad (1.21)$$

and causality is not broken.

For

$$\frac{\hbar\omega}{mc^2} > 1; \qquad \frac{\hbar\omega}{\alpha^2 mc^2} > 1 \qquad (1.22)$$

we obtain from formula (1.18)

$$\upsilon = \alpha c, \quad \upsilon < c. \qquad (1.23)$$

Considering formulae (1.21) and (1.23) we conclude that the hyperbolic diffusion equation (1.16) describes the thermal phenomena in accordance with special relativity theory and causality is not broken independently of laser beam energy.

When the amplitude of the laser beam approaches the critical electric field of quantum electrodynamics (Schwinger field [1.9]) the vacuum becomes polarized and electron – positron pairs are created in vacuum [1.9]. On a distance equal to the Compton length, $\lambdabar_C = \hbar/m_e c$, the work of critical field on an electron is equal to the electron rest mass energy $m_e c^2$, i.e. $eE_{Sch}\lambdabar_C = m_e c^2$. The dimensionless parameter

$$\frac{E}{E_{Sch}} = \frac{e\hbar E}{m_e^2 c^3} \qquad (1.24)$$

becomes equal to unity for electromagnetic wave intensity of the order of

$$I = \frac{c}{r_e \lambdabar_C^2} \frac{m_e c^2}{4\pi} \cong 4.7 \cdot 10^{29} \tfrac{W}{cm^2} \qquad (1.25)$$

where $r_e$ is the classical electron radius [1.10]. For such ultra high intensities the effects of nonlinear quantum electrodynamics plays a key role: laser beams excite virtual electron – positron pairs. As a result the vacuum acquires a finite electric and magnetic susceptibility which lead to the scattering of light by light. The cross section for the photon – photon interaction is given by:

$$\sigma_{\gamma\gamma \to \gamma\gamma} = \frac{973}{10125} \frac{\alpha^3}{\pi^2} r_e^2 \left(\frac{\hbar\omega}{m_e c^2}\right)^6, \qquad (1.26)$$

for $\hbar\omega/m_e c^2 < 1$ and reaches its maximum, $\sigma_{max} \approx 10^{-20} cm^2$ for $\hbar\omega \approx m_e c^2$ [1.9].

Considering formulae (1.25) and (1.26) we conclude that linear hyperbolic diffusion equation is valid only for the laser intensities $I \leq 10^{29}$ W/cm$^2$. For high intensities the nonlinear hyperbolic diffusion equation must be formulated and solved.

Table 1.1. Hierarchical structure of the thermal excitation

| Interaction | $\alpha$ | $mc^2\alpha$ |
|---|---|---|
| Electromagnetic | $1/137$ | $0.5/137$ |
| Strong | $\dfrac{15}{100}$ | $\dfrac{140 \cdot 15}{100}$ for pions |
|  |  | $\dfrac{1000 \cdot 15}{100}$ for nucleons |
| Quark - Quark | 1 | 417[*] |

* D.H. Perkins, Introduction to high energy physics, Addison – Wesley, USA 1987

References to Chapter 1

# Chapter 2
# Non-Fourier description of laser - matter interaction

## 2.1 Fundamentals of laser pulses interaction with matter

In this paragraph we will study the ultra-short thermal processes in the framework of the hyperbolic diffusion equation.

When an ultrafast thermal pulse (e.g. femtosecond pulse) interacts with a metal surface, the excited electrons become the main carriers of the thermal energy. For a femtosecond thermal pulse, the duration of the pulse is of the same order as the electron relaxation time. In this case, the hyperbolicity of the thermal energy transfer plays an important role [2.1].

Radiation deposition of energy in materials is a fundamental phenomenon to laser processing. It converts radiation energy into material's internal energy, which initiates many thermal phenomena, such as heat pulse propagation, melting and evaporation. The operation of many laser techniques requires an accurate understanding and control of the energy deposition and transport processes. Recently, radiation deposition and the subsequent energy transport in metals have been investigated with picosecond and femtosecond resolutions. Results show that during high-power and short-pulse laser heating, free electrons can be heated to an effective temperature much higher than the lattice temperature, which in turn leads to both a much faster energy propagation process and a much smaller lattice-temperature rise than those predicted from the conventional radiation heating model. Corkum et al. [2.2] found that this electron-lattice nonequilibrium heating mechanism can significantly increase the resistance of molybdenum and copper mirrors to thermal damage during high-power laser irradiation when the laser pulse duration is shorter than one nanosecond. Clemens et al. [2.3] studied thermal transport in multilayer metals during picosecond laser heating. The measured temperature response in the first 20 ps was found to be different from predictions of the conventional Fourier model. Due to the relatively low temporal resolution of the experiment (~ 4 ps), however, it is difficult to determine whether this difference is the result of nonequilibrium laser heating or is due to other heat conduction mechanisms, such as non-Fourier heat conduction, or reflection and refraction of thermal waves at interfaces. Heat is conducted in solids through electrons and phonons. In metals, electrons dominate the heat conduction, while in insulators and semiconductors, phonons are the major heat carriers. Table 2.1 lists important features of the electrons and phonons.

Table 2.1. General Features of Heat Carriers

|  | Free Electron | Phonon |
|---|---|---|
| Generation | ionization or excitation | lattice vibration |
| Propagation media | vacuum or media | media only |
| Statistics | Fermion | Boson |
| Dispersion | $E = \hbar^2 q^2/(2m)$ | $E = E(q)$ |
| Velocity (m·s$^{-1}$) | ~ $10^6$ | ~ $10^3$ |

The traditional thermal science, or macroscale heat transfer, employs phenomenological laws, such as Fourier's law, without considering the detailed motion of the heat carriers. Decreasing dimensions, however, have brought an increasing need for understanding the heat transfer processes from the microscopic point of view of the heat carriers. The response of the electron and phonon gases to the external perturbation initiated by laser irradiation can be described with the help of a memory function of the system. To that aim, let us consider the generalized Fourier law [2.1]:

$$q(t) = -\int_{-\infty}^{t} K(t-t')\nabla T(t')dt', \qquad (2.1)$$

where $q(t)$ is the density of a thermal energy flux, $T(t')$ is the temperature of electrons and $K(t - t')$ is a memory function for thermal processes. The density of thermal energy flux satisfies the following equation of heat conduction:

$$\frac{\partial}{\partial t}T(t) = \frac{1}{\rho c_V}\nabla^2 \int_{-\infty}^{t} K(t-t')T(t')dt', \qquad (2.2)$$

where $\rho$ is the density of charge carriers and $c_V$ is the specific heat of electrons in a constant volume. We introduce the following equation for the memory function describing the Fermi gas of charge carriers:

$$K(t-t') = K_1 \lim_{t_0 \to 0} \delta(t-t'-t_0). \tag{2.3}$$

In this case, the electron has a very "short" memory due to thermal disturbances of the state of equilibrium. Combining Eqs. (2.3) and (2.2) we obtain

$$\frac{\partial}{\partial t} T(t) = \frac{1}{\rho c_V} K_1 \nabla^2 T. \tag{2.4}$$

Equation (2.4) has the form of the parabolic equation for heat conduction (PHC). Using this analogy, Eq. (2.4) may be transformed as follows:

$$\frac{\partial}{\partial t} T(t) = D_T \nabla^2 T. \tag{2.5}$$

where the heat diffusion coefficient $D_T$ is defined as follows:

$$D_T = \frac{K_1}{\rho c_V}. \tag{2.6}$$

From Eq. (2.6), we obtain the relation between the memory function and the diffusion coefficient

$$K(t-t') = D_T \rho c_V \lim_{t_0 \to 0} \delta(t-t'-t_0). \tag{2.7}$$

In the case when the electron gas shows a "long" memory due to thermal disturbances, one obtains for memory function

$$K(t-t') = K_2 \tag{2.8}$$

When Eq. (2.8) is substituted to the Eq. (2.2) we obtain

$$\frac{\partial}{\partial t} T = \frac{K_2}{\rho c_v} \nabla^2 \int_{-\infty}^{t} T(t')dt, \tag{2.9}$$

Differentiating both sides of Eq. (2.9) with respect to $t$, we obtain

$$\frac{\partial^2 T}{\partial t^2} = \frac{K_2}{\rho c_V} \nabla^2 T. \tag{2.10}$$

Equation (2.10) is the hyperbolic wave equation describing thermal wave propagation in a charge carrier gas in a metal film. Using a well-known form of the wave equation,

$$\frac{1}{v^2}\frac{\partial^2 T}{\partial t^2} = \nabla^2 T. \tag{2.11}$$

and comparing Eqs. (2.10) and (2.11), we obtain the following form for the memory function:

$$K(t-t') = \rho c_V v^2 \tag{2.12}$$

$$v = \text{finite}, \ v < \infty.$$

As the third case, "intermediate memory" will be considered:

$$K(t-t') = \frac{K_3}{\tau}\exp\left[-\frac{(t-t')}{\tau}\right], \tag{2.13}$$

where $\tau$ is the relaxation time of thermal processes. Combining Eqs. (2.13) and (2.2) we obtain

$$c_V\frac{\partial^2 T}{\partial t^2} + \frac{c_V}{\tau}\frac{\partial T}{\partial t} = \frac{K_3}{\rho\tau}\nabla^2 T \tag{2.14}$$

and

$$K_3 = D_T c_V \rho. \tag{2.15}$$

Thus, finally,

$$\frac{\partial^2 T}{\partial t^2} + \frac{1}{\tau}\frac{\partial T}{\partial t} = \frac{D_T}{\tau}\nabla^2 T. \tag{2.16}$$

Equation (2.16) is the hyperbolic equation for heat conduction (HHC), in which the electron gas is treated as a Fermion gas. The diffusion coefficient $D_T$ can be written in the form [2.1]

$$D_T = \frac{1}{3}v_F^3\tau, \tag{2.17}$$

where $v_F$ is the Fermi velocity for the electron gas in a semiconductor. Applying Eq. (2.17) we can transform the hyperbolic equation for heat conduction, Eq.(2.16), as follows:

$$\frac{\partial^2 T}{\partial t^2} + \frac{1}{\tau}\frac{\partial T}{\partial t} = \frac{1}{3}v_F^3 \nabla^2 T. \tag{2.18}$$

Let us denote the velocity of disturbance propagation in the electron gas as *s*:

$$s = \sqrt{\frac{1}{3}} v_F. \tag{2.19}$$

Using the definition of *s*, Eq. (2.18) may be written in the form

$$\frac{1}{s^2}\frac{\partial^2 T}{\partial t^2} + \frac{1}{\tau s^2}\frac{\partial T}{\partial t} = \nabla^2 T. \tag{2.20}$$

For the electron gas, treated as the Fermi gas, the velocity of sound propagation is described by the equation (2.20)

$$v_s = \left(\frac{p_F^2}{3mm^*}\left(1+F_0^S\right)\right)^{1/2}, \quad p_F = mv_F, \tag{2.21}$$

where *m* is the mass of a free (non-interacting) electron and *m\** is the effective electron mass. Constant $F_0^S$ represents the magnitude of carrier-carrier interaction in the Fermi gas. In the case of a very weak interaction, $m \to m^*$ and $F_0^S \to 0$, so according to Eq. (2.21),

$$v_S = \frac{mv_F}{\sqrt{3}m} = \sqrt{\frac{1}{3}} v_F. \tag{2.22}$$

To sum up, we can make a statement that for the case of weak electron-electron interaction, sound velocity $v_S = \sqrt{1/3}v_F$ and this velocity is equal to the velocity of thermal disturbance propagation *s*. From this we conclude that the hyperbolic equation for heat conduction Eq. (2.20), is identical as the equation for second sound propagation in the electron gas:

$$\frac{1}{v_S^2}\frac{\partial^2 T}{\partial t^2} + \frac{1}{\tau v_S^2}\frac{\partial T}{\partial t} = \nabla^2 T. \tag{2.23}$$

Using the definition expressed by Eq. (2.17) for the heat diffusion coefficient, Eq. (2.23) may be written in the form

$$\frac{1}{v_S^2}\frac{\partial^2 T}{\partial t^2} + \frac{1}{D_T}\frac{\partial T}{\partial t} = \nabla^2 T. \tag{2.24}$$

The mathematical analysis of Eq. (2.23) leads to the following conclusions:

1. In the case when $v_S^2 \to \infty$, $\tau v_S^2$ is finite, Eq. (2.26) transforms into the parabolic equation for heat diffusion:

$$\frac{1}{D_T}\frac{\partial T}{\partial t} = \nabla^2 T. \tag{2.25}$$

2. In the case when $\tau \to \infty$, $v_S$ is finite, Eq. (2.24) transforms into the wave equation:

$$\frac{1}{v_S^2}\frac{\partial^2 T}{\partial t^2} = \nabla^2 T. \tag{2.26}$$

Equation (2.26) describes propagation of the thermal wave in the electron gas. From the point of view of theoretical physics, condition $v_S \to \infty$ violates the special theory of relativity. From this theory we know that there is a limited velocity of interaction propagation and this velocity $v_{lim} = c$, where $c$ is the velocity of light in a vacuum.

Multiplying both sides of Eq. (2.24) by $c^2$, we obtain

$$\frac{c^2}{v_S^2}\frac{\partial^2 T}{\partial t^2} + \frac{c^2}{D_T}\frac{\partial T}{\partial t} = c^2 \nabla^2 T, \tag{2.27}$$

Denoting $\beta = v_S/c$, Eq. (2.27) may be written in the form

$$\frac{1}{\beta^2}\frac{\partial^2 T}{\partial t^2} + \frac{1}{\widetilde{D}_T}\frac{\partial T}{\partial t} = c^2 \nabla^2 T, \tag{2.28}$$

where $\widetilde{D}_T = \tau \beta^2$, $\beta < 1$. On the basis of the above considerations, we conclude that the heat conduction equation, which satisfies the special theory of relativity, acquires the form of the partial hyperbolic Eq. (2.28). The rejection of the first component in Eq. (2.28) violates the special theory of relativity.

In one-dimensional flow of heat in metals, the hyperbolic heat transport equation is given by (2.18).

$$\tau\frac{\partial^2 T}{\partial t^2} + \frac{\partial T}{\partial t} = D_T \frac{\partial^2 T}{\partial x^2}, \qquad D_T = \frac{1}{3}v_F^2 \tau, \tag{2.29}$$

where $\tau$ denotes the relaxation time, $D_T$ is the diffusion coefficient and $T$ is the temperature. Introducing the non-dimensional spatial coordinate $z = x/\lambdabar$, where $\lambdabar = \lambda/2\pi$ denotes the reduced mean free path, Eq. (2.29) can be written in the form:

$$\frac{1}{v'^2}\frac{\partial^2 T}{\partial t^2} + \frac{2a}{v'^2}\frac{\partial T}{\partial t} = \frac{\partial^2 T}{\partial z^2}, \tag{2.30}$$

where

$$v' = \frac{v}{\lambda} \qquad a = \frac{1}{2\tau} \tag{2.31}$$

In Eq. (2.31) $v$ denotes the velocity of heat propagation [2.1], $v = (D/\tau)^{1/2}$.

In the paper by C. De Witt-Morette and See Kit Fong [2.4], the path-integral solution of Eq. (2.30) was obtained. It was shown, that for the initial condition of the form:

$$T(z,0) = \Phi(z) \qquad \text{an "arbitrary" function}$$

$$\frac{\partial T(z,t)}{\partial t}\bigg|_{t=0} = 0 \tag{2.32}$$

the general solution of the Eq. (2.29) has the form:

$$T(z,t) = \frac{1}{2}[\Phi(z,t) + \Phi(z,-t)]e^{-at}$$

$$+ \frac{a}{2}e^{-at}\int_0^t d\eta [\Phi(z,\eta) + \Phi(z,-\eta)] \tag{2.33}$$

$$+ \left[ I_0(a(t^2-\eta^2)^{1/2}) + \frac{t}{(t^2-\eta^2)^{1/2}} I_1(a(t^2-\eta^2)^{1/2}) \right]$$

In Eq. (2.33), $I_0(x)$ and $I_1(x)$ denote the modified Bessel function of zero and first order respectively.

Let us consider the propagation of the initial thermal wave with velocity $v'$, i.e.,

$$\Phi(z - v't) = \sin(z - v't) \tag{2.34}$$

In that case, the integral in (2.33) can be computed analytically, $\Phi(z,t) + \Phi(z,-t) = 2\sin z \cos(v't)$ and the integrals on the right-hand side of (2.33) can be done explicitly [2.4]; we obtain:

$$F(z,t) = e^{-at}\left[\frac{a}{w_1}\sin(w_1 t) + \cos(w_1 t)\right]\sin z, \qquad v' \geq a \qquad (2.35)$$

and

$$F(z,t) = e^{-at}\left[\frac{a}{w_2}\sinh(w_2 t) + \cosh(w_2 t)\right]\sin z, \qquad v' < a \qquad (2.36)$$

where $w_1 = (v'^2 - a^2)^{1/2}$ and $w_2 = (a^2 - v'^2)^{1/2}$.

In order to clarify the physical meaning of the solutions given by formulas (2.35) and (2.36), we observe that $v' = v/\lambda$ and $w_1$ and $w_2$ can be written as:

$$v_1 = \lambda w_1 = v\left(1 - \left(\frac{1}{2\tau\omega}\right)^2\right)^{1/2}, \qquad 2\tau\omega > 1$$

$$v_2 = \lambda w_2 = v\left(\left(\frac{1}{2\tau\omega}\right)^2 - 1\right)^{1/2}, \qquad 2\tau\omega < 1 \qquad (2.37)$$

where $\omega$ denotes the pulsation of the initial thermal wave. From formula (2.37), it can be concluded that we can define the new effective thermal wave velocities $v_1$ and $v_2$. Considering formulas (2.36) and (2.37), we observe that the thermal wave with velocity $v_2$ is very quickly attenuated in time. It occurs that when $\omega^{-1} > 2\tau$, the scatterings of the heat carriers diminish the thermal wave.

It is interesting to observe that in the limit of a very short relaxation time, i.e., when $\tau \to 0$, $v_2 \to \infty$, because for $\tau \to 0$ Eq. (2.29) is the Fourier parabolic equation.

It can be concluded, that for $\omega^{-1} > 2\tau$, the Fourier equation is relevant equation for the description of the thermal phenomena in metals. For $\omega^{-1} > 2\tau$, the scatterings are slower than in the preceding case and attenuation of the thermal wave is weaker. In that case, $\tau \neq 0$ and $v_1$ is always finite:

$$v_1 = v\left(1 - \left(\frac{1}{2\tau\omega}\right)^2\right)^{1/2} < v \qquad (2.38)$$

For $\tau \to 0$, i.e., for very rare scatterings $v_1 \to v$ and Eq. (2.29) is a nearly free thermal wave equation. For $\tau$ finite the $v_1 < v$ and thermal wave propagates in the medium with smaller velocity than the velocity of the initial thermal wave.

Considering the formula (2.35), one can define the change of the phase of the initial thermal wave $\beta$, i.e.:

$$\tan[\beta] = \frac{a}{w_1} = \frac{1}{2\tau\omega}\frac{1}{\sqrt{1-\frac{1}{4\tau^2\omega^2}}}, \qquad 2\tau\omega > 1 \qquad (2.39)$$

We conclude that the scatterings produce the change of the phase of the initial thermal wave. For $\tau \to \infty$ (very rare scatterings), $\tan[\beta] = 0$.

## 2.2. High-order wave equation for thermal transport phenomena

In papers [2.5, 2.6] the generalized thermal wave equation (with the third order time derivative) was considered.

For heat transport initiated by ultrashort laser pulses, when $\Delta t > \tau$ one obtains the second order PDE for quantum thermal phenomena

$$\tau\frac{\partial^2 T}{\partial t^2} + \frac{\partial T}{\partial t} = D\nabla^2 T - \frac{2V}{Dm}T, \qquad D = \frac{\hbar}{m} \qquad (2.40)$$

Equation (2.40) can be written as

$$\frac{2V\tau}{Dm}T + \tau\frac{\partial T}{\partial t} + \tau^2\frac{\partial^2 T}{\partial t^2} = \tau D\nabla^2 T. \qquad (2.41)$$

For $V = 0$ equation (2.41) can be written as:

$$\tau\frac{\partial T}{\partial t} + \tau^2\frac{\partial^2 T}{\partial t^2} = D\tau\nabla^2 T. \qquad (2.42)$$

From dimensional analysis we can write new equation:

$$\left(\tau\frac{\partial}{\partial t} + \tau^2\frac{\partial^2}{\partial t^2} + \tau^3\frac{\partial^3}{\partial t^3}\right)T = D\tau\nabla^2 T. \qquad (2.43)$$

Let us write Eq. (2.43) in the form

$$\kappa\nabla^2 T = \varepsilon\frac{\partial^2 T}{\partial t^2} + \mu\frac{\partial T}{\partial t} + \mu_3\frac{\partial^3 T}{\partial t^3} \qquad (2.44)$$

where

$$\kappa = D\tau \qquad \varepsilon = \tau^2, \qquad \mu = \tau, \qquad \mu_3 = \tau^3 \qquad (2.45)$$

Equation (2.44) yields the characteristic polynomial equation

$$p(s,jk) = \mu_3 s^3 + \varepsilon s^2 + \mu s + \kappa k^2 = 0 \tag{2.46}$$

Equation (2.44) was investigated, for oscillating transport phenomena, by P.M. Ruiz. It was shown in paper [2.7] that one of the solution of Eq. (2.44) is nonphysical i.e. increasing with time.

In one-dimensional case one obtains from Eq. (2.43)

$$\tau \frac{\partial T}{\partial t} + \tau^2 \frac{\partial^2 T}{\partial t^2} + \tau^3 \frac{\partial^3 T}{\partial t^3} = \kappa \frac{\partial^2 T}{\partial x^2} \tag{2.47}$$

The second-order hyperbolic equation

$$\tau \frac{\partial T}{\partial t} + \tau^2 \frac{\partial^2 T}{\partial t^2} = \kappa \frac{\partial^2 T}{\partial x^2} \tag{2.48}$$

was discussed in monograph [2.1]. Below we analyze the third-order wave equation

$$\tau^2 \frac{\partial^3 T}{\partial t^3} = D \frac{\partial^2 T}{\partial x^2} \tag{2.49}$$

in the case of thermal processes induced by attosecond laser pulses

$$\tau = \frac{D}{v^2}, \qquad v = \alpha c, \tag{2.50}$$

and equation (2.49) can be rewritten as

$$\frac{\partial^2 T}{\partial x^2} = \beta \frac{\partial^3 T}{\partial t^3}, \qquad \beta = \frac{\hbar}{mv^4} = \frac{D}{v^4}. \tag{2.51}$$

We seek a solution of equation (2.51) of the form

$$T(x,t) = A e^{i(kx - \omega t)}. \tag{2.52}$$

Substituting equation (2.52) to Eq. (2.51) one obtains

$$(ik)^2 = \beta(-i\omega)^3. \tag{2.53}$$

This shows that equation (2.52) is the solution of the third-order PDE i.e. Eq. (2.51) is the third-order wave equation if

$$\beta = \frac{(ik)^2}{(-i\omega)^3} = \frac{\hbar}{mv^2} = \frac{D}{v^2} \tag{2.54}$$

where *v* is the speed of propagation of thermal energy [2.1]. Substituting Eq. (2.54) to Eq. (2.52) one obtains

$$T(x,t) = Ae^{i\left[\frac{x}{\sqrt{2}\lambda} - \omega t\right]} e^{-\frac{1}{\sqrt{2}}\frac{x}{\lambda}} + Be^{-i\left[\frac{x}{\sqrt{2}\lambda} - \omega t\right]} e^{\frac{1}{\sqrt{2}}\frac{x}{\lambda}} \qquad (2.55)$$

where $\lambda$ is mean free path.

The second term in Eq. (2.55) tends to infinity for $x/\lambda \gg 1$ and is to be omitted. The final solution of Eq. (2.52) has the form

$$T(x,t) = e^{-\frac{1}{\sqrt{2}}\frac{x}{\lambda}} Ae^{i\left[\frac{x}{\sqrt{2}\lambda} - \omega t\right]} \qquad (2.56)$$

and describes the strongly damped thermal wave.

It is interesting to observe that for electromagnetic interaction the third-order time derivative $d^3x/dt^3$ also describes the damping of the electron motion due to the self interaction of the charges [2.8].

In this paragraph the third-order wave equation for thermal processes was derived and solved. It was shown that the additional third-order derivative term describes the strongly damped wave. For $x/\lambda > 1$, where $\lambda$ is the mean free path, the influence of third-order derivative term $\tau^3 \partial^3 T/\partial t^3$ can be omitted when describing the thermal phenomena induced by ultrashort laser pulses.

## 2.3. Quantum heat transport equation

Dynamical processes are commonly investigated using laser pump-probe experiments with a pump pulse exciting the system of interest and a second probe pulse tracking is temporal evolution. As the time resolution attainable in such experiments depends on the temporal definition of the laser pulse, pulse compression to the attosecond domain is a recent promising development.

After the standards of time and space were defined the laws of classical physics relating such parameters as distance, time, velocity, temperature are assumed to be independent of accuracy with which these parameters can be measured. It should be noted that this assumption does not enter explicitly into the formulation of classical physics. It implies that together with the assumption of existence of an object and really independently of any measurements (in classical physics) it was tacitly assumed that *there was a possibility of an unlimited increase in accuracy of measurements.* Bearing in mind the "atomicity" of time i.e. considering the smallest time period, the Planck time, the above statement is obviously not true. Attosecond laser pulses we are at the limit of laser time resolution.

With attosecond laser pulses belong to a new Nano – World where size becomes comparable to atomic dimensions, where transport phenomena follow different laws from that in the macro world. This first stage of miniaturization, from $10^{-3}$ m to $10^{-6}$ m is over and the new one, from $10^{-6}$ m to $10^{-9}$ m just beginning. The Nano – World is a quantum world with all the predicable and non-predicable (yet) features.

In this paragraph, we develop and solve the quantum relativistic heat transport equation for nanoscale transport phenomena where external forces exist [2.9]. In paragraph 2.1 we developed the new hyperbolic heat transport equation which generalizes the Fourier heat transport equation for the rapid thermal processes. The hyperbolic heat transport equation (HHT) for the fermionic system has be written in the form (2.23)

$$\frac{1}{\left(\frac{1}{3}v_F^2\right)}\frac{\partial^2 T}{\partial t^2} + \frac{1}{\tau\left(\frac{1}{3}v_F^2\right)}\frac{\partial T}{\partial t} = \nabla^2 T \ , \qquad (2.57)$$

where $T$ denotes the temperature, $\tau$ the relaxation time for the thermal disturbance of the fermionic system, and $v_F$ is the Fermi velocity.

In what follows we develop the new formulation of the HHT, considering the details of the two fermionic systems: electron gas in metals and the nucleon gas.

For the electron gas in metals, the Fermi energy has the form

$$E_F^e = (3\pi)^2 \frac{n^{2/3}\hbar^2}{2m_e}, \qquad (2.58)$$

where $n$ denotes the density and $m_e$ electron mass. Considering that

$$n^{-1/3} \sim a_B \sim \frac{\hbar^2}{me^2}, \qquad (2.59)$$

and $a_B$ = Bohr radius, one obtains

$$E_F^e \sim \frac{n^{2/3}\hbar^2}{2m_e} \sim \frac{\hbar^2}{ma^2} \sim \alpha^2 m_e c^2, \qquad (2.60)$$

where $c$ = light velocity and $\alpha = 1/137$ is the fine-structure constant for electromagnetic interaction. For the Fermi momentum $p_F$ we have

$$p_F^e \sim \frac{\hbar}{a_B} \sim \alpha m_e c, \qquad (2.61)$$

and, for Fermi velocity $v_F$,

$$v_F^e \sim \frac{p_F}{m_e} \sim \alpha c. \qquad (2.62)$$

Formula (2.62) gives the theoretical background for the result presented in paragraph 2.1. Considering formula (2.62), equation HHT can be written as

$$\frac{1}{c^2}\frac{\partial^2 T}{\partial t^2} + \frac{1}{c^2\tau}\frac{\partial T}{\partial t} = \frac{\alpha^2}{3}\nabla^2 T. \qquad (2.63)$$

As seen from (2.63), the HHT equation is a relativistic equation, since it takes into account the finite velocity of light.

For the nucleon gas, Fermi energy equals

$$E_F^N = \frac{(9\pi)^{2/3}\hbar^2}{8mr_0^2}, \qquad (2.64)$$

where $m$ denotes the nucleon mass and $r_0$, which describes the range of strong interaction, is given by

$$r_0 = \frac{\hbar}{m_\pi c}, \qquad (2.65)$$

wherein $m_\pi$ is the pion mass. From formula (2.65), one obtains for the nucleon Fermi energy

$$E_F^N \sim \left(\frac{m_\pi}{m}\right)^2 mc^2. \qquad (2.66)$$

In analogy to the Eq. (2.60), formula (2.66) can be written as

$$E_F^N \sim \alpha_s^2 mc^2, \qquad (2.67)$$

where $\alpha_s = \frac{m_\pi}{m} \cong 0.15$ is the fine-structure constant for strong interactions. Analogously, we obtain the nucleon Fermi momentum

$$p_F^e \sim \frac{\hbar}{r_0} \sim \alpha_s mc \tag{2.68}$$

and the nucleon Fermi velocity

$$v_F^N \sim \frac{pF}{m} \sim \alpha_s c, \tag{2.69}$$

and HHT for nucleon gas can be written as

$$\frac{1}{c^2}\frac{\partial^2 T}{\partial t^2} + \frac{1}{c^2 \tau}\frac{\partial T}{\partial t} = \frac{\alpha_s^2}{3}\nabla^2 T. \tag{2.70}$$

In the following, the procedure for the discretization of temperature $T(\vec{r},t)$ in hot fermion gas will be developed. First of all, we introduce the reduced de Broglie wavelength

$$\begin{aligned}\lambda_B^e &= \frac{\hbar}{m_e v_h^e}, & v_h^e &= \frac{1}{\sqrt{3}}\alpha c, \\ \lambda_B^N &= \frac{\hbar}{m v_h^N}, & v_h^N &= \frac{1}{\sqrt{3}}\alpha_s c,\end{aligned} \tag{2.71}$$

and the mean free paths $\lambda_e$ and $\lambda_N$

$$\lambda^e = v_h^e \tau^e, \qquad \lambda^N = v_h^N \tau^N. \tag{2.72}$$

In view of formulas (2.71) and (2.72), we obtain the HHC for electron and nucleon gases

$$\frac{\lambda_B^e}{v_h^e}\frac{\partial^2 T}{\partial t^2} + \frac{\lambda_B^e}{\lambda^e}\frac{\partial T}{\partial t} = \frac{\hbar}{m_e}\nabla^2 T^e, \tag{2.73}$$

$$\frac{\lambda_B^N}{v_h^N}\frac{\partial^2 T}{\partial t^2} + \frac{\lambda_B^N}{\lambda^N}\frac{\partial T}{\partial t} = \frac{\hbar}{m}\nabla^2 T^N. \tag{2.74}$$

Equations (2.73) and (2.74) are the hyperbolic partial differential equations which are the master equations for heat propagation in Fermi electron and nucleon gases. In the following, we will study the quantum limit of heat transport in the fermionic systems. We define the quantum heat transport limit as follows:

$$\lambda^e = \lambdabar_B^e, \qquad \lambda^N = \lambdabar_B^N. \tag{2.75}$$

In that case, Eqs. (2.73) and (2.74) have the form

$$\tau^e \frac{\partial^2 T^e}{\partial t^2} + \frac{\partial T^e}{\partial t} = \frac{\hbar}{m_e}\nabla^2 T^e, \qquad (2.76)$$

$$\tau^N \frac{\partial^2 T^N}{\partial t^2} + \frac{\partial T^N}{\partial t} = \frac{\hbar}{m}\nabla^2 T^N, \qquad (2.77)$$

where

$$\tau^e = \frac{\hbar}{m_e (v_h^e)^2}, \qquad \tau^N = \frac{\hbar}{m (v_h^N)^2}. \qquad (2.78)$$

Equations (2.76) and (2.77) define the master equation for quantum heat transport (QHT). Having the relaxation times $\tau^e$ and $\tau^N$, one can define the "pulsations" $\omega_h^e$ and $\omega_h^N$

$$\omega_h^e = (\tau^e)^{-1}, \qquad \omega_h^N = (\tau^N)^{-1}, \qquad (2.79)$$

or

$$\omega_h^e = \frac{m_e (v_h^e)^2}{\hbar}, \qquad \omega_h^N = \frac{m (v_h^N)^2}{\hbar},$$

i.e.,

$$\omega_h^e \hbar = m_e (v_h^e)^2 = \frac{m_e \alpha^2}{3} c^2,$$
$$\omega_h^N \hbar = m (v_h^N)^2 = \frac{m \alpha_s^2}{3} c^2. \qquad (2.80)$$

The formulas (2.80) define the Planck-Einstein relation for heat quanta $E_h^e$ and $E_h^N$

$$E_h^e = \omega_h^e \hbar = m_e (v_h^e)^2,$$
$$E_h^N = \omega_h^N \hbar = m_N (v_h^N)^2. \qquad (2.81)$$

The heat quantum with energy $E_h = \hbar\omega$ can be named the *heaton*, in complete analogy to the *phonon, magnon, roton*, etc. For $\tau^e, \tau^N \to 0$, Eqs. (2.76) and (2.80) are the Fourier equations with quantum diffusion coefficients $D^e$ and $D^N$

$$\frac{\partial T^e}{\partial t} = D^e \nabla^2 T^e, \qquad D^e = \frac{\hbar}{m_e}, \qquad (2.82)$$

$$\frac{\partial T^N}{\partial t} = D^N \nabla^2 T^N, \qquad D^N = \frac{\hbar}{m}. \tag{2.83}$$

For finite $\tau^e$ and $\tau^N$, for $\Delta t < \tau^e$, $\Delta t < \tau^N$, Eqs. (2.76) and (2.77) can be written as

$$\frac{1}{(v_h^e)^2}\frac{\partial^2 T^e}{\partial t^2} = \nabla^2 T^e, \tag{2.84}$$

$$\frac{1}{(v_h^N)^2}\frac{\partial^2 T^N}{\partial t^2} = \nabla^2 T^N. \tag{2.85}$$

Equations (2.84) and (2.85) are the wave equations for quantum heat transport (QHT). For $\Delta t > \tau$, one obtains the Fourier equations (2.82) and (2.83).

In what follows, the dimensionless form of the QHT will be used. Introducing the reduced time $t'$ and reduced length $x'$,

$$t' = t/\tau, \qquad x' = \frac{x}{v_h \tau}, \tag{2.86}$$

one obtains, for QHT,

$$\frac{\partial^2 T^e}{\partial t^2} + \frac{\partial T^e}{\partial t} = \nabla^2 T^e, \tag{2.87}$$

$$\frac{\partial^2 T^N}{\partial t^2} + \frac{\partial T^N}{\partial t} = \nabla^2 T^N. \tag{2.88}$$

and, for QFT,

$$\frac{\partial T^e}{\partial t} = \nabla^2 T^e, \tag{2.89}$$

$$\frac{\partial T^N}{\partial t} = \nabla^2 T^N. \tag{2.90}$$

## 2.4. *Proca* thermal equation

It is quite interesting that the *Proca* type equation can be obtained for thermal phenomena. In the following starting with the hyperbolic heat diffusion equation the *Proca* equation for thermal processes will be developed and solved [2.9].

In paper [2.9] the relativistic hyperbolic transport equation was developed:

$$\frac{1}{v^2}\frac{\partial^2 T}{\partial t^2} + \frac{m_0 \gamma}{\hbar}\frac{\partial T}{\partial t} = \nabla^2 T. \tag{2.91}$$

In equation (2.91) $v$ is the velocity of heat waves, $m_0$ is the mass of heat carrier and $\gamma$ – the Lorentz factor, $\gamma = \left(1 - \frac{v^2}{c^2}\right)^{-1/2}$. As was shown in paper [2.9] the heat energy (*heaton temperature*) $T_h$ can be defined as follows:

$$T_h = m_0 \gamma v^2. \tag{2.92}$$

Considering that $v$, the thermal wave velocity equals [2.9]

$$v = \alpha c \tag{2.93}$$

where $\alpha$ is the coupling constant for the interactions which generate the *thermal wave* ($\alpha = 1/137$ and $\alpha = 0.15$ for electromagnetic and strong forces respectively). The *heaton temperature* is equal to

$$T_h = \frac{m_0 \alpha^2 c^2}{\sqrt{1 - \alpha^2}}. \tag{2.94}$$

Based on equation (2.94) one concludes that the *heaton temperature* is a linear function of the mass $m_0$ of the heat carrier. It is interesting to observe that the proportionality of $T_h$ and the heat carrier mass $m_0$ was observed for the first time in ultrahigh energy heavy ion reactions measured at CERN [2.10]. In paper [2.10] it was shown that the temperature of pions, kaons and protons produced in Pb+Pb, S+S reactions are proportional to the mass of particles. Recently, at Rutherford Appleton Laboratory (RAL), the VULCAN LASER was used to produce the elementary particles: electrons and pions [2.11].

In this chapter the forced relativistic heat transport equation will be studied and solved. In the case of the forced heat transport the master equation is of the form [2.1]:

$$\frac{1}{v^2}\frac{\partial^2 T}{\partial t^2} + \frac{m}{\hbar}\frac{\partial T}{\partial t} + \frac{2Vm}{\hbar^2}T - \nabla^2 T = 0. \tag{2.95}$$

The relativistic generalization of equation (2.95) is quite obvious:

$$\frac{1}{v^2}\frac{\partial^2 T}{\partial t^2} + \frac{m_0 \gamma}{\hbar}\frac{\partial T}{\partial t} + \frac{2V m_0 \gamma}{\hbar^2}T - \nabla^2 T = 0. \tag{2.96}$$

It is worthwhile noting that in order to obtain a non-relativistic equation we put $\gamma = 1$.

When the external force is present $F(x,t)$ the forced damped heat transport is obtained instead of equation (2.96) (in one dimensional case):

$$\frac{1}{v^2}\frac{\partial^2 T}{\partial t^2} + \frac{m_0\gamma}{\hbar}\frac{\partial T}{\partial t} + \frac{2Vm_0\gamma}{\hbar^2}T - \frac{\partial^2 T}{\partial x^2} = F(x,t). \tag{2.97}$$

The hyperbolic relativistic quantum heat transport equation, (2.97), describes the forced motion of heat carriers which undergo scattering ($\frac{m_0\gamma}{\hbar}\frac{\partial T}{\partial t}$ term) and are influenced by the potential term ($\frac{2Vm_0\gamma}{\hbar^2}T$).

Equation (2.97) is the *Proca* thermal equation and can be written as [2.9]:

$$\left(\Box + \frac{2Vm_0\gamma}{\hbar^2}\right)T + \frac{m_0\gamma}{\hbar}\frac{\partial T}{\partial t} = F(x,t),$$

$$\Box = \frac{1}{v^2}\frac{\partial^2}{\partial t^2} - \frac{\partial^2}{\partial x^2}. \tag{2.98}$$

We seek the solution of equation (2.98) in the form

$$T(x,t) = e^{-t/2\tau}u(x,t) \tag{2.99}$$

where $\tau = \hbar/(mv^2)$ is the relaxation time. After substituting equation (2.99) in equation (2.98) we obtain a new equation

$$(\Box + q)u(x,t) = e^{t/2\tau}F(x,t) \tag{2.100}$$

and

$$q = \frac{2Vm}{\hbar^2} - \left(\frac{mv}{2\hbar}\right)^2 \tag{2.101}$$

$$m = m_0\gamma \tag{2.102}$$

In free space i.e. when $F(x,t) \to 0$ equation (2.100) reduces to

$$(\Box + q)u(x,t) = 0 \tag{2.103}$$

which is essentially the free *Proca* equation.

The *Proca* equation describes the interaction of the laser pulse with the matter. As was shown in paper [2.1] the quantization of the temperature field leads to the *heatons* – quanta of thermal energy with a mass $m_h = \hbar/\tau v_h^2$ [2.1], where $\tau$ is the relaxation time and $v_h$ is the finite velocity for heat propagation. For $v_h \to \infty$, i.e. for $c \to \infty$, $m_o \to 0$. it can be concluded that in non-relativistic approximation ($c$ = infinite) the *Proca* equation is the diffusion equation for massless photons and heatons.

### 3.3. Solution of the *Proca* thermal equation

For the initial *Cauchy* condition:

$$u(x,0) = f(x), \qquad u_t(x,0) = g(x) \qquad (2.104)$$

the solution of the *Proca* equation has the form (for $q > 0$) [2.9]

$$\begin{aligned}
u(x,t) &= \frac{f(x-vt) + f(x+vt)}{2} \\
&+ \frac{1}{2v} \int_{x-vt}^{x+vt} g(\varsigma) J_0\left[\sqrt{q(v^2 t^2 - (x-\varsigma)^2)}\right] d\varsigma \\
&- \frac{\sqrt{q} vt}{2} \int_{x-vt}^{x+vt} f(\varsigma) \frac{J_1\left[\sqrt{q(v^2 t^2 - (x-\varsigma)^2)}\right]}{\sqrt{v^2 t^2 - (x-\varsigma)^2}} d\varsigma \\
&+ \frac{1}{2v} \int_0^t \int_{x-v(t-t')}^{x+v(t-t')} G(\varsigma,t') J_0\left[\sqrt{q(v^2 (t-t')^2 - (x-\varsigma)^2)}\right] dt' d\varsigma.
\end{aligned} \qquad (2.105)$$

where $G = e^{t/2\tau} F(x,t)$.

When $q < 0$ solution of *Proca* equation has the form:

$$\begin{aligned}
u(x,t) &= \frac{f(x-vt) + f(x+vt)}{2} \\
&+ \frac{1}{2v} \int_{x-vt}^{x+vt} g(\varsigma) I_0\left[\sqrt{-q(v^2 t^2 - (x-\varsigma)^2)}\right] d\varsigma \\
&- \frac{\sqrt{-q} vt}{2} \int_{x-vt}^{x+vt} f(\varsigma) \frac{I_1\left[\sqrt{-q(v^2 t^2 - (x-\varsigma)^2)}\right]}{\sqrt{v^2 t^2 - (x-\varsigma)^2}} d\varsigma \\
&+ \frac{1}{2v} \int_0^t \int_{x-v(t-t')}^{x+v(t-t')} G(\varsigma,t') I_0\left[\sqrt{-q(v^2 (t-t')^2 - (x-\varsigma)^2)}\right] dt' d\varsigma.
\end{aligned} \qquad (2.106)$$

When $q = 0$ equation (2.100) is the forced thermal equation

$$\frac{1}{v^2}\frac{\partial^2 u}{\partial t^2} - \frac{\partial^2 u}{\partial x^2} = G(x,t). \tag{2.107}$$

On the other hand one can say that equation (2.107) is distortion-less hyperbolic equation. The condition $q = 0$ can be rewritten as:

$$V\tau = \frac{\hbar}{8} \tag{2.108}$$

The equation (2.108) is the analogous to the Heisenberg uncertainty relation. Considering equation (2.92) equation (2.108) can be written as:

$$V = \frac{T_h}{8}, \qquad V < T_h. \tag{2.109}$$

It can be stated that distortion-less waves can be generated only if $T_h > V$. For $T_h < V$, i.e. when the "Heisenberg rule" is broken, the shape of the thermal waves is changed.

In this chapter we developed the relativistic thermal transport equation for an attosecond laser pulse interaction with matter. It is shown that the equation obtained is the *Proca* equation, well known in relativistic electrodynamics for massive photons. As the *heatons* are massive particles the analogy is well founded. Considering that for an attosecond laser pulse the damped term in Eq. (2.108) tends to 1, the transport phenomena are well described by the *Proca* equation.

# Chapter 3
# Time delayed processes in biophysics

## 3.1. Clusters and aggregates of atoms

Clusters and aggregates of atoms in the nanometer range (currently called nanoparticles) are systems intermediate in several respects, between simple molecules and bulk materials and have been the subject of intensive work.

In this paragraph, we investigate the thermal relaxation phenomena in nanoparticles – microtubules within the frame of the quantum heat transport equation. In reference [3.1], the thermal inertia of materials, heated with laser pulses faster than the characteristic relaxation time was investigated. It was shown, that in the case of the ultra-short laser pulses it was necessary to use the hyperbolic heat conduction (HHC). For microtubules the diameters are of the order of the de Broglie wave length. In that case quantum heat transport must be used [3.1] to describe the transport phenomena,

$$\tau \frac{\partial^2 T}{\partial t^2} + \frac{\partial T}{\partial t} = \frac{\hbar}{m}\nabla^2 T, \tag{3.1}$$

where $T$ denotes the temperature of the heat carrier, and $m$ denotes its mass and $\tau$ is the relaxation time. The relaxation time $\tau$ is defined as [3.1]

$$\tau = \frac{\hbar}{mv_h^2}, \tag{3.2}$$

where $v_h$ is the thermal pulse propagation rate

$$v_h = \frac{1}{\sqrt{3}}\alpha c \tag{3.3}$$

In equation (3.3) $\alpha$ is a coupling constant (for the electromagnetic interaction $\alpha = e^2/\hbar c$ and $c$ denotes the speed of light in vacuum. Both parameters $\tau$ and $v_h$ characterizes completely the thermal energy transport on the atomic scale and can be termed *"atomic relaxation time"* and *"atomic"* heat diffusivity.

Both τ and $v_h$ contain constants of Nature, $\alpha$, $c$. Moreover, on an atomic scale there is no shorter time period than and smaller velocity than that build from of constants in Nature. Consequently, one can call τ and $v_h$ the *elementary relaxation time* and *elementary diffusivity,* which characterizes heat transport in the elementary building block of matter, the atom. In the following, starting with elementary τ and $v_h$, we shall describe thermal relaxation processes in microtubules which consist of the $N$ components (molecules) each with elementary τ and $v_h$. With this in view, we use the Pauli-Heisenberg inequality [3.1]

$$\Delta r \Delta p \geq N^{\frac{1}{3}} \hbar, \tag{3.4}$$

where $r$ denotes the characteristic dimension of the nanoparticle and $p$ is the momentum of the energy carriers. The Pauli-Heisenberg inequality expresses the basic property of the $N$ – fermionic system. In fact, compared to the standard Heisenberg inequality

$$\Delta r \Delta p \geq \hbar, \tag{3.5}$$

we observe that, in this case the presence of the large number of identical fermions forces the system either to become spatially more extended for a fixed typical momentum dispension, or to increase its typical momentum dispension for a fixed typical spatial extension. We could also say that for a fermionic system in its ground state, the average energy per particle increases with the density of the system.

An illustrative means of interpreting the Pauli-Heisenberg inequality is to compare Eq. (3.4) with Eq. (3.5) and to think of the quantity on the right hand side of it as the *effective fermionic Planck constant*

$$\hbar^f(N) = N^{\frac{1}{3}} \hbar. \tag{3.6}$$

We could also say that antisymmetrization, which typifies fermionic amplitudes amplifies those quantum effects which are affected by the Heisenberg inequality.

Based on equation (3.6), we can recalculate the relaxation time $\tau$, equation (3.2) and the thermal speed $v_h$, equation (3.3) for a nanoparticle consisting of $N$ fermions

$$\hbar \leftarrow \hbar^f(N) = N^{\frac{1}{3}} \hbar \tag{3.7}$$

and obtain

$$v_h^f = \frac{e^2}{\hbar^f(N)} = \frac{1}{N^{\frac{1}{3}}} v_h,$$

(3.8)

$$\tau^f = \frac{\hbar^f}{m(v_h^f)^2} = N\tau.$$

(3.9)

The number *N* particles in a nanoparticle (sphere with radius *r)* can be calculated using the equation (we assume that density of a nanoparticle does not differ too much from that of the bulk material)

$$N = \frac{\frac{4\pi}{3} r^3 \rho A Z}{\mu}$$  (3.10)

and for non spherical shapes with semi axes *a, b, c*

$$N = \frac{\frac{4\pi}{3} abc \rho A Z}{\mu}$$  (3.11)

where $\rho$ is the density of the nanoparticle, *A* is the Avogardo number, $\mu$ is the molecular mass of theparticles in grams and *Z* is the number of valence electrons.

Using equations (3.8) and (3.9), we can calculate the de Broglie wave length $\lambda_B^f$ and mean free path $\lambda_{mfp}^f$ for nanoparticles

$$\lambda_B^f = \frac{\hbar^f}{m v_{th}^f} = N^{\frac{2}{3}} \lambda_B,$$  (3.12)

$$\lambda_{emfp}^f = v_{th}^f \tau_{th}^f = N^{\frac{2}{3}} \lambda_{mfp},$$  (3.13)

where $\lambda_B$ and $\lambda_{mfp}$ denote the de Broglie wave length and the mean free path for heat carriers in nanoparticles (e.g. microtubules).

## 3.2. Quantum transport in microtubules

In paper [3.2] we develop the Klein - Gordon thermal equation for microtubules. Microtubules are essential to cell functions. In neurons, microtubules help and regulate synaptic activity responsible for learning and cognitive function. Whereas microtubules have traditionally been considered to be purely structural elements, recent evidence has revealed that mechanical, chemical and electrical signaling and a communication function also exist as a result of the microtubule interaction with membrane structures by linking proteins, ions and voltage fields respectively. The characteristic dimensions of the microtubules; a crystalline cylinder 10 nm internal diameter, are of the order of the de Broglie length for electrons in atoms. When the characteristic length of the structure is of the order of the de Broglie wave length, then the signaling phenomena must be described by the quantum transport theory. n order to describe quantum transport phenomena in microtubules it is necessary to use equation (3.14) with the relaxation time described as follows:

$$\tau \frac{\partial^2 T}{\partial t^2} + \frac{\partial T}{\partial t} = \frac{\hbar}{m}\nabla^2 T,$$

$$\tau = \frac{2\hbar}{mv^2} = \frac{\hbar}{E}. \qquad (3.14)$$

where the relaxation time is the de-coherence time, i.e. the time before the wave function collapses, when the transition classical $\rightarrow$ quantum phenomena is considered.

In the following we consider the time $\tau$ for atomic and multiatomic phenomena. As was shown in Chapter 2 for atomic phenomena

$$\tau_a \approx 10^{-17}\,\text{s} \qquad (3.15)$$

and when we consider multiatomic transport phenomena, with $N$ equal number of aggregates involved the equation is (3.9)

$$\tau_N = N\tau_a \qquad (3.16)$$

The Penrose – Hameroff Orchestrated Objective Reduction Model (OrchOR) [3.3] proposes that quantum superposition – computation occurs in nanotube automata within brain neurons and glia. Tubulin subunits within microtubules act as qubits, switching between states on a nanosecond scale, governed by London forces in hydrophobic pockets. These oscillations are tuned and orchestrated by microtubule associated proteins (MAPs) providing a feedback loop between the biological system and the quantum state. These qubits interact computationally by non-local quantum entanglement, according to the Schrödinger equation with preconscious processing continuing until the threshold for objective reduction (OR) is reached $(E = \hbar/T)$. At that instant, collapse occurs, triggering a "moment of awareness" or a conscious event – an event that determines particular configurations of Planck scale experiential geometry and corresponding classical states of nanotubes automata that regulate synaptic and other neural functions.

A sequence of such events could provide a forward flow of subjective time and stream of consciousness. Quantum states in nanotubules may link to those in nanotubules in other neurons and glia by tunneling through gap functions,

Table 3.1. The de-coherence relaxation time

| Event | $T$ [ms] | $E$ | $N$ number of aggregates | $T$ [ms] |
|---|---|---|---|---|
| Buddhist moment of awareness nucleons | 13 | $4 \cdot 10^{15}$ | $10^{15}$ | 10 |
| Coherent 40 Hz oscillations | 25 | $2 \cdot 10^{15}$ | $10^{15}$ | 10 |
| EEG alpha rhytm (8 to 12 Hz) | 100 | $10^{14}$ | $10^{14}$ | 1 |
| Libet's sensory threshold | 100 | $10^{14}$ | $10^{14}$ | 1 |

permitting extension of the quantum state through significant volumes of the brain.

Based on $E = \hbar/T$, the size and extension of Orch OR events which correlate with a subjective or neurophysiological description of conscious events can be calculated. In Table 3.1 the calculated $T$ (Penrose-Hameroff) and $\tau$ – equation (3.16) are presented [3.2].

## 3.3. Heisenberg uncertainty principle for thermal phenomena in microtubules

We shall now develop the generalized quantum heat transport equation for microtubules which also includes the potential term. Thus, we are able to use the analogy of the Schrödinger and quantum heat transport equations. If we consider, for the moment, the parabolic heat transport equation with the second derivative term omitted

$$\frac{\partial T}{\partial t} = \frac{\hbar}{m} \nabla^2 T. \qquad (3.17)$$

If the real time $t \to it/2$, $T \to \Psi$, Eq. (3.17) has the form of a free Schrödinger equation

$$i\hbar \frac{\partial \Psi}{\partial t} = -\frac{\hbar^2}{2m} \nabla^2 \Psi. \tag{3.18}$$

The complete Schrödinger equation has the form

$$i\hbar \frac{\partial \Psi}{\partial t} = -\frac{\hbar^2}{2m} \nabla^2 \Psi + V\Psi, \tag{3.19}$$

where $V$ denotes the potential energy. When we go back to real time $t \to 2it$, $\Psi \to T$, the new parabolic heat transport is obtained

$$\frac{\partial T}{\partial t} = \frac{\hbar}{m} \nabla^2 T - \frac{2V}{\hbar} T. \tag{3.20}$$

Equation (3.20) describes the quantum heat transport for $\Delta t > \tau$. For heat transport initiated by ultra-short laser pulses, when $\Delta t < \tau$ one obtains the generalized quantum hyperbolic heat transport equation

$$\tau \frac{\partial^2 T}{\partial t^2} + \frac{\partial T}{\partial t} = \frac{\hbar}{m} \nabla^2 T - \frac{2V}{\hbar} T. \tag{3.21}$$

Considering that $\tau = \hbar/mv^2$, Eq. (3.21) can be written as follows:

$$\frac{1}{v^2} \frac{\partial^2 T}{\partial t^2} + \frac{m}{\hbar} \frac{\partial T}{\partial t} + \frac{2Vm}{\hbar^2} T = \nabla^2 T. \tag{3.22}$$

Equation (3.22) describes the heat flow when apart from the temperature gradient, the potential energy $V$ (is present.)

In the following, we consider one-dimensional heat transfer phenomena, i.e

$$\frac{1}{v^2} \frac{\partial^2 T}{\partial t^2} + \frac{m}{\hbar} \frac{\partial T}{\partial t} + \frac{2Vm}{\hbar^2} T = \frac{\partial^2 T}{\partial x^2}. \tag{3.23}$$

We seek a solution in the form

$$T(x,t) = e^{\frac{1}{2\tau}} u(x,t). \tag{3.24}$$

for the quantum heat transport equation (3.23)

After substitution of Eq. (3.24) into Eq. (3.23), one obtains

$$\frac{1}{v^2} \frac{\partial^2 u}{\partial t^2} - \frac{\partial^2 u}{\partial x^2} + qu(x,t) = 0. \tag{3.25}$$

where

$$q = \frac{2Vm}{\hbar^2} - \left(\frac{mv}{2\hbar}\right)^2 \qquad (3.26)$$

In the following, we consider a constant potential energy $V = V_0$. The general solution of Eq. (3.25) for the Cauchy boundary conditions,

$$u(x,0) = f(x), \qquad \left[\frac{\partial u(x,t)}{\partial t}\right]_{t=0} = F(x), \qquad (3.27)$$

has the form [3.1]

$$u(x,t) = \frac{f(x-vt) + f(x+vt)}{2} + \frac{1}{2v}\int_{x-vt}^{x+vt} \Phi(x,y,z)dz, \qquad (3.28)$$

where

$$\Phi(x,t,z) = \frac{1}{v}J_0\left(\frac{b}{v}\sqrt{(z-x)^2 - v^2t^2}\right) + btf(z)\frac{J_0\left(\frac{b}{v}\sqrt{(z-x)^2 - v^2t^2}\right)}{\sqrt{(z-x)^2 - v^2t^2}},$$

$$b = \left(\frac{mv^2}{2\hbar}\right) - \frac{2Vm}{\hbar^2}v^2 \qquad (3.29)$$

and $J_0(z)$ denotes the Bessel function of the first kind. Considering equations (3.24), (3.25), (3.26) the solution of Eq. (3.23) describes the propagation of the distorted thermal quantum waves with characteristic lines $x = \pm vt$. We can define the distortionless thermal wave as the wave which preserves the shape in the potential energy $V_0$ field. The condition for conserving the shape can be expressed as

$$q = \frac{2Vm}{\hbar^2} - \left(\frac{mv}{2\hbar}\right)^2 \qquad (3.30)$$

When Eq. (3.30) holds, Eq. (3.31) has the form

$$\frac{\partial^2 u(x,t)}{\partial t^2} = v^2 \frac{\partial^2 u}{\partial x^2}. \qquad (3.31)$$

Equation (3.31) is the quantum wave equation with the solution (for Cauchy boundary conditions (3.27))

$$u(x,t) = \frac{f(x-vt) + f(x+vt)}{2} + \frac{1}{2v}\int_{x-vt}^{x+vt} F(z)dz. \tag{3.32}$$

It is interesting to observe, that condition (3.30) has an analog in classical theory of the electrical transmission line. In the context of the transmission of an electromagnetic field, the condition $q = 0$ describes the Heaviside distortionless line. Eq. (3.30) – the distortionless condition – can be written as

$$V_0 \tau \approx \hbar, \tag{3.33}$$

We can conclude, that in the presence of the potential energy $V_0$ one can observe the undisturbed quantum thermal wave in microtubules only when *the Heisenberg uncertainty relation for thermal processes* (3.33) is fulfilled.

The generalized quantum heat transport equation (GQHT) (3.23) leads to generalized Schrödinger equation for microtubules. After the substitution $t \to it/2$, $T \to \Psi$ in Eq. (3.23), one obtains the generalized Schrödinger equation (GSE)

$$i\hbar \frac{\partial \Psi}{\partial t} = -\frac{\hbar^2}{2m}\nabla^2\Psi + V\Psi - 2\tau\hbar \frac{\partial^2 \Psi}{\partial t^2}. \tag{3.34}$$

Considering that $\tau = \hbar/mv^2 = \hbar/m\alpha^2 c^2$ ($\alpha = 1/137$) is the fine-structure constant for electromagnetic interactions) Eq. (3.34) can be written as

$$i\hbar \frac{\partial \Psi}{\partial t} = -\frac{\hbar^2}{2m}\nabla^2\Psi + V\Psi - \frac{2\hbar^2}{m\alpha^2 c^2}\frac{\partial^2 \Psi}{\partial t^2}. \tag{3.35}$$

One can conclude, that for a time period $\nabla t < \hbar/m\alpha^2 c^2 \approx 10^{-17}$ s the description of quantum phenomena needs some revision. On the other hand, for $\nabla t > 10^{-17}$ in GSE the second derivative term can be omitted and as a result the Schrödinger equation SE is obtained, i.e.

$$i\hbar \frac{\partial \Psi}{\partial t} = -\frac{\hbar^2}{2m}\nabla^2\Psi + V\Psi \tag{3.36}$$

It is interesting to observe, that GSE was discussed also in the context of the sub-quantal phenomena.

In conclusion a study of the interactions of the attosecond laser pulses with matter can shed light on the applicability of the SE in a study of ultra-short sub-quantal phenomena.

The structure of Eq. (3.25) depends on the sign of the parameter $q$. For quantum heat transport phenomena with electrons as the heat carriers the parameter $q$ is a function of the potential barrier height $V_0$ and velocity $v$.

The initial Cauchy condition

$$u(x,0) = f(x), \qquad \frac{\partial u(x,0)}{\partial t} = g(x), \qquad (3.37)$$

and the solution of the Eq. (3.25) has the form

$$u(x,t) = \frac{f(x-vt) + f(x+vt)}{2}$$
$$+ \frac{1}{2v} \int_{x-vt}^{x+vt} g(\varsigma) I_0\left[\sqrt{-q(v^2 t^2 - (x-\varsigma)^2)}\right] d\varsigma \qquad (3.38)$$
$$- \frac{\left(v\sqrt{-q}\right)t}{2} \int_{x-vt}^{x+vt} f(\varsigma) \frac{I_1\left[\sqrt{-q(v^2 t^2 - (x-\varsigma)^2)}\right]}{\sqrt{v^2 t^2 - (x-\varsigma)^2}} d\varsigma.$$

When $q > 0$ Eq. (3.25) is the *Klein – Gordon equation* (K-G), which is well known from applications in elementary particle and nuclear physics.

For the initial Cauchy condition (3.37), the solution of the (K-G) equation can be written as

$$u(x,t) = \frac{f(x-vt) + f(x+vt)}{2}$$
$$+ \frac{1}{2v} \int_{x-vt}^{x+vt} g(\varsigma) J_0\left[\sqrt{q(v^2 t^2 - (x-\varsigma)^2)}\right] d\varsigma \qquad (3.39)$$
$$- \frac{\left(v\sqrt{q}\right)t}{2} \int_{x-vt}^{x+vt} f(\varsigma) \frac{J_0'\left[\sqrt{-q(v^2 t^2 - (x-\varsigma)^2)}\right]}{\sqrt{v^2 t^2 - (x-\varsigma)^2}} d\varsigma.$$

Both solutions (3.38) and (3.39) exhibit the domains of dependence and influence on the *modified Klein-Gordon* and *Klein-Gordon equation*. These domains, which characterize the maximum speed at which a thermal disturbance can travel are determined by the principal terms of the given equation (i.e., the second derivative terms) and do not depend on the lower order terms. It can be concluded that these equations and the wave equation (for $m = 0$) have identical domains of dependence and influence.

References to Chapter 3

[3.1] Kozlowski, M. and Marciak-Kozlowska, J. *From Quarks to Bulk Matter,*

Hadronic Press, 2001.

[3.2] J. Marciak - Kozlowska, M. Kozlowski, M. Pelc, Lasers in Engineering *16* (2006) 195.

[3.3] Hameroff, S. R. and Tuszyński, J. (1996). Conscious events as orchestrated spacetime selections, *Journal of Consciousness Studies, 3*, 36.

# Chapter 4
# Time delayed processes in archeology

## 4.1. Introduction

　　The origins of European agriculture are normally sought in the Near East. The earliest indicators of agriculture in the form of cultivation of cereals and pulses are rearing of animals come from Zagros foothills. Their age, 11700 – 8400 BC corresponds to the cool dry climatic period followed by a rapid increase in rainfall at the beginning of the Holocene (~ 10000 BC).

　　During the early stages of agricultural development (Preceramic Neolithic 9800 – 7500 BC) the rapid increase in the number of sites is noticeable in both the foothills and the surrounding plains accompanied by the appearance of large settlements with complicated masonry structures and fortifications (e.g. Jericho).

　　At a later stage the core area of early agricultural settlements shifts to the north to the eastern highlands and inner depressions of Asia Minor. The most outstanding case of early agricultural development in this area is Catal Hüyük, a Neolithic town on the Konya Plain (6500 – 5700 BC).

　　The earliest sites with developed agricultural economies in Europe, dated 6400 – 6000 BC are found in the intermontane depressions of Greece (Thessaly, Beotia and Pelloponesse). Genetic features of the cultigens and the general character of the material culture leave no doubt to their Near – Eastern origins. Significantly, the early Neolithic sites in the Marmara Sea are of a more recent age (6100 – 5600 BC), being culturally distinct from the Early Neolithic in Greece. This implies that the Neolithic communities could penetrate the Balkan Peninsula from Western Asia by means of navigation.

The Neolithic spread further, plausibly via the Strouma axis in the northeast and the Vardar – Morava axis in the north. The ensuing development saw a rapid growth of Neolithic settlements in the depressions of northern Thrace the Lower and Middle Danube catchment basin (5900 – 5500 BC).

The next stage in the Neolithic development saw the emergence of new sites on the Tisza Plain in 5600 – 5500 BC [4.1].

The new sites later spread over the vast areas of the loess plains of Central Europe, mostly along the Danube, Rhine – Mainz and Vistula axes. This spread occurred within the range of 5600 – 4800 BC within the most probable age of 5154 ± 62 BC [4.2].

Judging from the number of sites the population in the Near East started increasing ~ 15000 BC. Ammerman and Cavalli – Sforza [4.3] focused on measuring the rate of spread of early farming in Europe and derived the rate of spread $v$ ~ 1 km/year on average in Europe. The Danube and Rhine valleys the propagation paths had an increased propagation speed as did the Mediterranean west [4.4]. The speeds of propagation of the wave front $v$ in these areas are as follows:

$v$ = 1 km/year on average in Europe

$v$ = 4 – 6 km/year for the Danube – Rhine valleys

$v$ = 10 km/year for Mediterranean regions

Interpretations of these observations are usually based on the reaction – diffusion equation of population dynamics [4.5, 4.6]. Fort and Méndez [4.6, 4.7] discuss the front propagation rate resulting from the generalization of Fisher model, but their results are restricted to the homogenous system.

The aim of this work is to formulate and develop a model for the spread of incipient farming in Europe taking into account the environmental influences on the migration processes.

## 4.2. The model

Fick diffusion equation is a special case of the parabolic transport equation in which speed of perturbation propagation is infinite. Parabolic transport equation has been applied to the spread of advantageous genes [4.8] dispersion of biological population [4.9], epidemic models [4.10].

However if perturbation propagates at finite speed, Fick law does not hold [4.11]. This unphysical feature can be avoided by making use of hyperbolic transport equation [4.11]. The hyperbolic propagation equation, HPE has been very recently applied to the spread of epidemics [4.12], forest fire models [4.13] and chemical system [4.14].

An interesting application of the Fick law to the migration in neolithic Europe was presented in [4.15].

Such a model as was presented in [4.15] provides a consistent explanation for the origin of Indo – European languages [4.16] and also finds remarkable support from the observed gene frequencies [4.17].

In this paper we develop the model for the population migration following the method presented in [4.11].

Let $n(x,t)$ stands for population density (measured in number of families per square kilometer) where $x$ is Cartesian coordinate and $t$ is time. We assume that a well defined time scale, $\tau$ between two successive migration steps exists. The migrated population interacts with environment. This interaction we will model by potential $V$. in that case the hyperbolic migration equation can be written as [4.11]:

$$\tau \frac{\partial^2 p}{\partial t^2} + \frac{\partial p}{\partial t} + \frac{V p_k}{D E_k} p = D \frac{\partial^2 p}{\partial x^2} + G(x,t) \qquad (4.1)$$

In equation (4.1) $D$ is the migration diffusion coefficient, $p_k$ and $E_k$ denote the momentum and kinetic energy of migration and $G(x,t)$ is the population growth.

Equation (4.1) is the Heaviside equation for population density [4.4]. It is the hyperbolic equation which describes the migration with memory. Memory term $\tau \partial^2 p / \partial t^2$ changes the type of the equation. For $\tau = 0$ equation [4.1] is the parabolic Fick equation for population migration.

We seek solution of the Eq.(4.1) in the form

$$p(x,t) = e^{-1/2\tau} u(x,t) \qquad (4.2)$$

After substituting of Eq. (4.2) to Eq.(4.1) one obtains

$$\frac{1}{v^2} \frac{\partial^2 u}{\partial t^2} - \frac{\partial^2 u}{\partial x^2} + u(x,t) \left[ -\frac{1}{4\tau D} + \frac{V p_k^2}{D^2 E} \right] = \frac{G(x,t)}{D} e^{1/2\tau}. \qquad (4.3)$$

For $V < 0$, i.e. for attractive potential equation (4.3) is the modified Klein – Gordon equation, i.e. so called "telegrapher equation". For $V > 0$ we have two possibilities:

$$\tau > \frac{ED}{4Vp_k^2} \qquad (4.4)$$

and

$$\tau < \frac{ED}{4Vp_k^2} \qquad (4.5)$$

In the case described by the inequality (4.4) equation (4.3) is the Klein – Gordon equation and for (4.5) Eq. (4.3) is the telegrapher equation.

In the following we will consider the $V < 0$, i.e. attractive potential; in that case equation (4.3) can be written as

$$\frac{1}{v^2}\frac{\partial^2 u}{\partial t^2} - \frac{\partial^2 u}{\partial x^2} - qu(x,t) = F(x,t). \qquad (4.6)$$

$$q = \frac{1}{4\tau D} + \frac{|V|p_k^2}{D^2 E_k}, \qquad q > 0 \qquad (4.7)$$

and

$$F(x,t) = \frac{G(x,t)e^{1/2\tau}}{D} \qquad (4.8)$$

Let us consider the initial condition

$$u(x,t) = f(x), \qquad u_t(x,0) = g(x) \qquad (4.9)$$
$$-\infty < x < \infty$$

for the equation (4.6). Then the general solution of Eq. (4.6) can be written as [4.18]:

$$u(x,t) = \frac{f(x-vt) + f(x+vt)}{2}$$
$$+ \frac{1}{2v}\int_{x-vt}^{x+vt} g(\varsigma) I_0\left[\left(\frac{q}{v}\right)^{1/2}\sqrt{v^2t^2 - (x-\varsigma)^2}\right]d\varsigma$$
$$- \frac{(qv)^{1/2}}{2} t \int_{x-vt}^{x+vt} f(\varsigma) \frac{I_1\left[\left(\frac{q}{v}\right)^{1/2}\sqrt{v^2t^2 - (x-\varsigma)^2}\right]}{\sqrt{v^2t^2 - (x-\varsigma)^2}}d\varsigma \qquad (4.10)$$
$$+ \frac{1}{2v}\int_0^t \int_{x-v(t-\eta)}^{x+v(t-\eta)} F(\varsigma,\eta) v I_0\left[\left(\frac{q}{v}\right)^{1/2}\sqrt{v^2(t-\eta)^2 - (x-\varsigma)^2}\right]d\varsigma d\eta.$$

Considering Eq. (4.10) and Eq. (4.2) the solution of Eq. (4.1) can be written as:

$$p(x,t) = e^{-1/2\tau} u(x,t) \qquad (4.11)$$

where the $u(x,t)$ is described by formula (4.10).

As can be seen from formula (4.11) the general solution of Eq. (4.1) is the sum of the wave motion described by the term

$$\frac{f(x-vt)+f(x+vt)}{2}$$

and the lag term.

In conclusion one can say that the migration can be described as the damped wave motion which for $t \to \infty$ is going to diffusion of the population.

References to Chapter 4

Acknowledgement

I thank very much Dr M.Kozlowski for the critical reading the manuscript and comments